\documentclass[a4paper]{article}

\usepackage{INTERSPEECH2022}

\title{Unsupervised domain adaptation for speech recognition with unsupervised error correction}
\name{Long Mai, Julie Carson-Berndsen}
\address{
  ML-Labs, School of Computer Science, University College Dublin, Ireland}
\email{long.mai@ucdconnect.ie, julie.berndsen@ucd.ie}

\begin{document}
\maketitle
\begin{abstract}
The transcription quality of automatic speech recognition (ASR) systems degrades significantly when transcribing audios coming from unseen domains. We propose an unsupervised error correction method for unsupervised ASR domain adaption, aiming to recover transcription errors caused by domain mismatch. Unlike existing correction methods that rely on transcribed audios for training, our approach requires only unlabeled data of the target domains in which a pseudo-labeling technique is applied to generate correction training samples. To reduce over-fitting to the pseudo data, we also propose an encoder-decoder correction model that can take into account additional information such as dialogue context and acoustic features. Experiment results show that our method obtains a significant word error rate (WER) reduction over non-adapted ASR systems. The correction model can also be applied on top of other adaptation approaches to bring an additional improvement of 10\% relatively.

\end{abstract}

\section{Introduction}
\label{sec:intro}

There is an increasing number of companies building speech-based applications ranging from voice transcription and translation to personal assistants and commercial chatbots. Due to the high cost of training a speech recognition model, companies often rely on off-the-shelf ASR systems such as Google ASR and Amazon Transcribe. As these systems are often trained on general labeled data, domain adaptation is a must to satisfy customer needs, e.g. one may want to adapt the American ASR model for the recognition of conversational accented speech. 

A simple approach for domain adaption is to fine-tune the pre-trained acoustic model on a sufficient amount of labeled data of the target domain. However, this introduces a high cost for the collection and annotation of the data. As a result, many methods have been proposed to perform domain adaption in a low-resource setting such as semi-supervised learning \cite{semi-low,semi-low2} and domain adversarial training \cite{semi-adv}.

The study presented in this paper focuses on unsupervised domain adaptation approaches, aiming to adapt the source ASR model to a target domain using only unlabeled data of that domain. 
The self-training method (or pseudo labeling) \cite{unsup-self} has recently demonstrated strong performances in unsupervised ASR domain adaptation. Standard self-training \cite{self-standard} proceeds by fine-tuning the ASR model on a dataset combining human-labeled data from the source domain with pseudo-labeled data from the target domain. The process can be repeated for multiple rounds until no significant improvement is observed. To avoid local minima after each round of training, data augmentation techniques and external language models are often integrated \cite{fb-iter}. Since the language model is trained independently on a separate text-corpus and is kept unchanged during the self-training process, they do not take into account the characteristic error distribution made by the ASR model \cite{gg-cor}. This leaves a potential for applying a correction model to recover these errors, and hence, leads to better pseudo transcriptions. Another limitation of current ASR domain adaptation approaches is that they often require full control over the source ASR system and its training data. This is not applicable for those who want to adapt off-the-shelf systems, such as Google ASR and Amazon Transcribe, as the users are not allowed to modify the source ASR model.

To over these limitations, this paper introduces a novel unsupervised domain adaptation method, aiming to adapt the ASR transcription output instead of the ASR model itself. We accomplish this by proposing an ASR error correction method to transform erroneous (out-of-domain) transcriptions into correct (in-domain) transcriptions. Similar to existing approaches, we employ a pre-trained encoder-decoder model as the backbone of the correction model. Instead of using ASR transcriptions as input and ground-truth transcriptions as output, we propose to use pseudo pairs of (sub-optimal ASR transcription, optimal ASR transcription) for training the model. These pseudo samples are generated by inferring the unlabeled audios with two ASR models that differ in terms of WER. To reduce over-fitting to the pseudo data, additional information, such as dialogue context and acoustic features, are also be taken into account alongside the 1-best ASR transcription. As the proposed method only modifies the ASR output, it can be applied in two settings:
\begin{enumerate}
    \item \textbf{Off-the-shelf.} In this case, there is no control over the ASR system and its training data. Therefore, standard adaptation methods adapting the ASR model cannot be applied. We show that our correction method can greatly reduce the WER by only modifying the ASR transcription. 
    \item \textbf{Self-training complementary.} In this case, full control over the ASR model is given. The self-training method is first applied to adapt the ASR model with unlabeled audios from the target domain. During inference, we place a newly trained correction model on top of the adapted ASR model to correct its predicted transcriptions. Experiment results show the correction model can bring an additional WER improvement of 10\% relatively. Although primarily focusing on the self-training technique, our framework is agnostic enough to be used in conjunction with other adaptation methods that modify the ASR model instead of its transcriptions. 
\end{enumerate}

\section{Related work}
\label{sec:related}

Domain mismatch between training and testing distributions degrades the performance of speech recognition systems significantly. A common approach to resolve this issue is adapting the original ASR model with labeled or unlabeled data of the unseen (target) domain. Many techniques have been proposed to make the adaptation process more accurate and efficient such as data augmentation \cite{unsup-aug}, distribution matching \cite{matching}, domain-adversarial training \cite{semi-adv2}, and self-training \cite{self2}. This study differentiates from the existing studies by proposing an error correction model for only adapting the ASR transcriptions.

ASR error correction, on the other hand, has been a very active research area in recent years. Prior studies often regard the task as sequence-to-sequence learning in which the corrupted transcription serves as input while the correct one serves as the output. Thus, the correction model can be implemented with architectures such as Long Short-Term Memory \cite{gg-cor} and Transformers \cite{nvidia-cor,adapt-cor}. However, it is difficult to detect and correct the errors using only the best ASR transcription. A recent study \cite{bart-cor} has shown that only 1/3 of the erroneous transcription can be corrected with 1-best transcription. As a result, many approaches have been proposed to take into account additional information such as contextual information \cite{uber-cor}, acoustic features \cite{cor_add1,cor_add2}, and visual features \cite{cor_add3}.

As existing correction models often share the same training data with the ASR system, they are often prone to over-fitting and might not be able to recover out-of-domain ASR errors when correcting samples coming from unseen domains. A simple solution \cite{adapt-cor} is to collect labeled data of the unseen (target) domain and then retrain the correction model to map out-of-domain ASR errors to in-domain terms using the ground truth transcriptions of the target domain.

Unlike all of the existing methods that require transcribed audios for training, our correction method relies only on unlabeled data, which is abundantly available. A new method to make better use of additional information, such as acoustic features and dialogue context, is also proposed.

\section{Proposed method}
\label{sec:method}
\subsection{Error correction for domain adaptation}

The goal of unsupervised domain adaptation is to adapt a source ASR model (trained on domain $X$) to target domain $Y$ using only unlabeled audios from that domain. This study focus on adapting the ASR transcriptions instead of the model itself. We accomplish this by using unlabeled data from domain $Y$ to train a correction model, which is then used to adapt the transcriptions of the source ASR model. Unlike existing methods that learn the mapping from erroneous transcriptions to ground-truth transcriptions, our approach learns the mapping from sub-optimal transcriptions to optimal transcriptions. The optimal one is not necessarily the ground-truth transcription but it should have a lower WER compared to the sub-optimal one. 

\begin{figure}[t]
  \centering
  {\includegraphics[width=\linewidth]{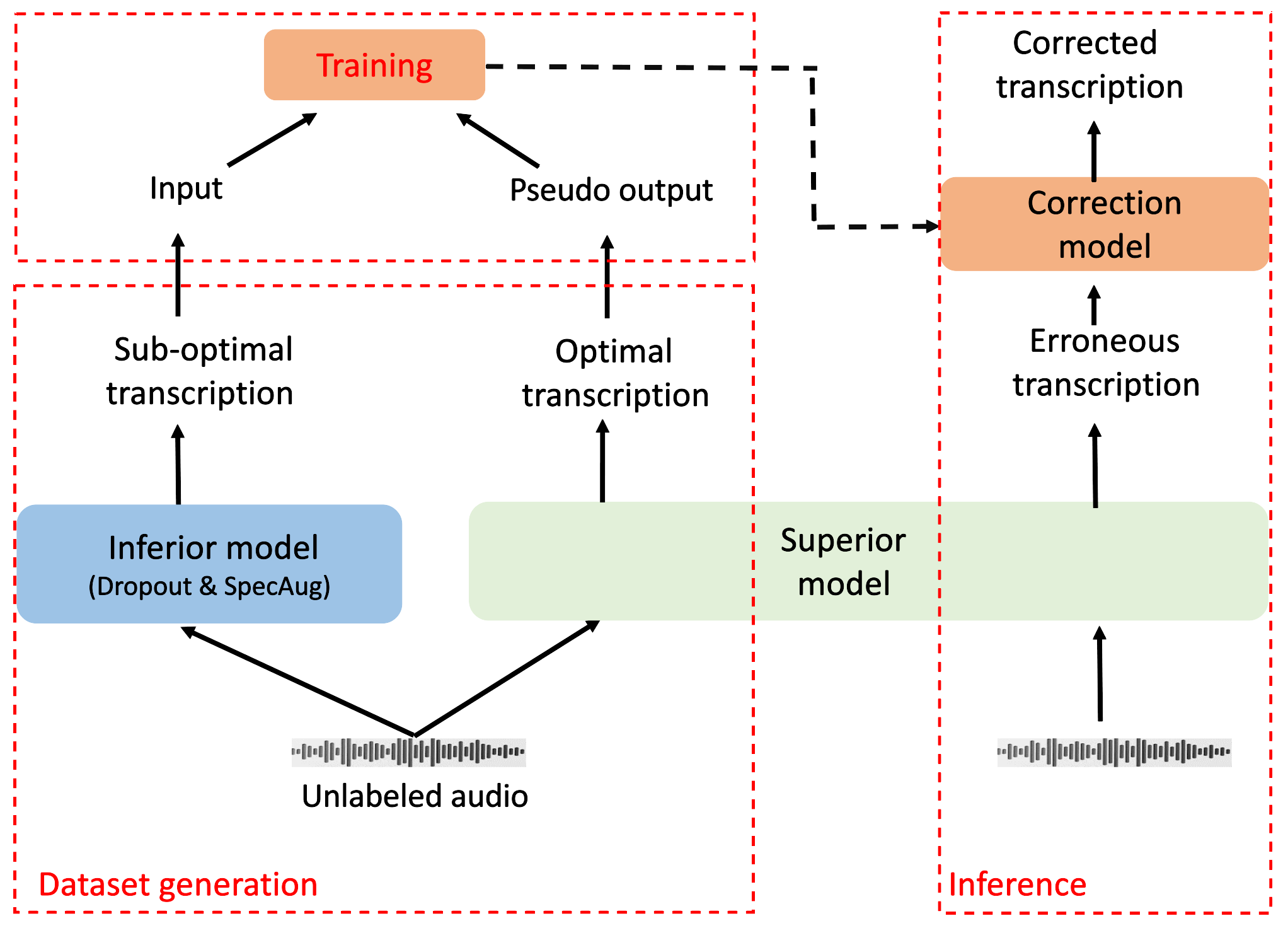}}
\caption{Unsupervised error correction framework}
\label{fig:framework}
\end{figure}

Figure \ref{fig:framework} illustrates how the correction is trained and used to make predictions. The first step is to construct a pseudo training dataset from unlabeled data. For each audio, a training pair of (sub-optimal ASR transcription, optimal ASR transcription) is generated by passing the same audio to an \textit{inferior} and a \textit{superior} ASR model respectively. The \textit{superior} is expected to perform significantly better than the \textit{inferior} in terms of WER. The next step is to train correction models using the pseudo dataset. At the prediction phase, new audio is first transcribed by the \textit{superior} model, i.e. the same model that was used for pseudo data generation. The resulting transcription, alongside other contextual information, is then fed into the trained correction model to produce the final result. Based on this framework, we can implement the two settings mentioned in Section \ref{sec:intro} as follows: 

\textbf{(1) For off-the-shelf setting}, the \textit{superior} model is the original (non-adapted) ASR model while the \textit{inferior} includes multiple poorly trained ASR models. The samples used to train these models are pseudo pairs of (unlabeled audio, optimal ASR transcription) generated by the \textit{superior}.

\textbf{(2) For self-training complementary setting}, the self-trained ASR model (final iteration) plays the role of the \textit{superior} model while the \textit{inferior} includes the non-adapted and the self-trained model at the first few iterations.

To ensure the \textit{superior} model performs better than the \textit{inferior} one, we can enable Dropout \cite{drop} and SpecAugment \cite{spec} during inference of the \textit{inferior} model, forcing it to generate sub-optimal transcriptions. In fact, we use these two techniques to not only control the WER of the pseudo correction dataset but also to generate more training samples.

\subsection{The correction model}

\begin{figure*}[t]
  \centering
  \includegraphics[width=12cm,height=4.5cm]{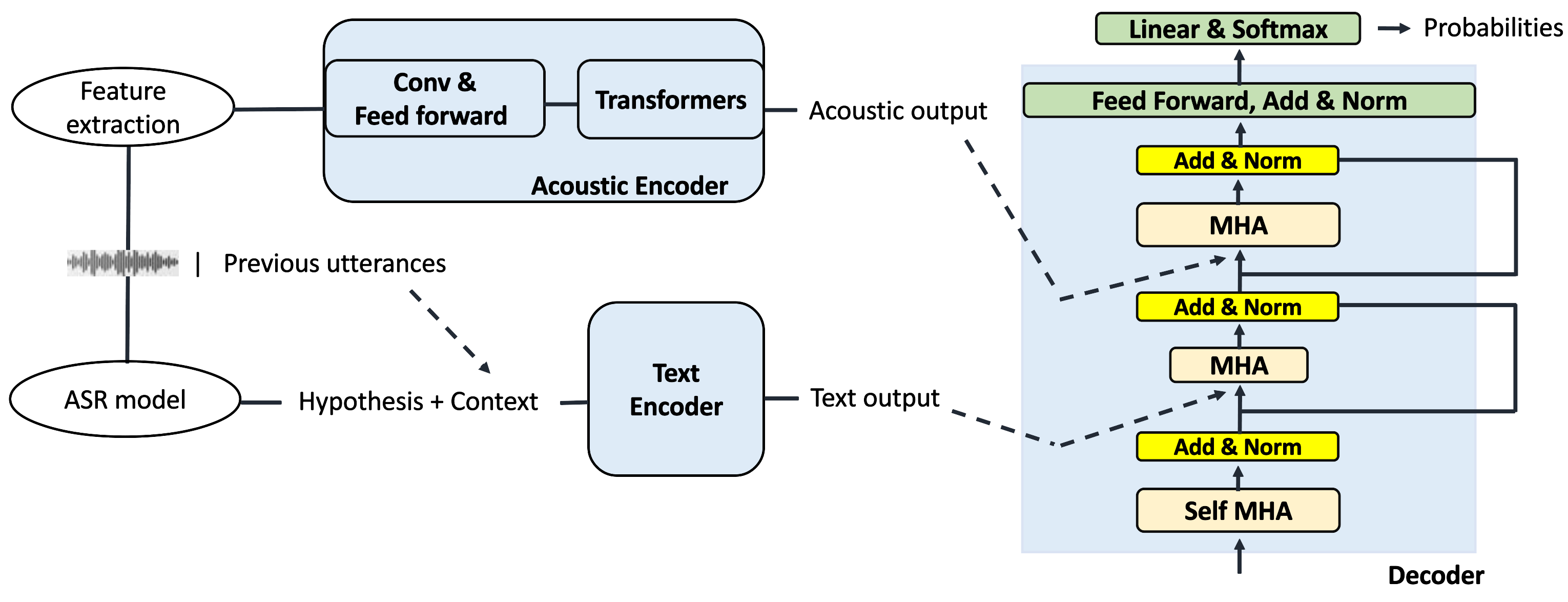}
\caption{Error correction model; \textit{MHA} denotes multi-headed attention}
\label{fig:model}
\end{figure*}

Our correction model is an extended version of a Transformers encoder-decoder architecture with sequential cross-attention mechanisms to incorporate multiple inputs. As illustrated in Figure \ref{fig:model}, the model consists of three components: a text encoder, an acoustic encoder, and a decoder.

The text encoder is a stack of 6 transformer layers in which each layer contains two sub-layers: self-attention and feed-forward. It receives the best hypothesis generated by the ASR model as input. To make use of contextual information, we simply append the dialogue context, i.e a concatenation of previous text utterances, at the end of the hypothesis. The encoder encodes the text input and produces text output \(H_{t} \in R^{L \times D}\), where $L$ is the number of tokens in the input, and $D = 768$ is the model hidden size.

The input of the acoustic encoder is a sequence of contextualized vectors \(H_{a} \in R^{T \times W}\) obtained by passing the audio through a pre-trained Wav2Vec2-Base model \cite{wav2vec}, where $T$ is the number of 20ms timesteps and $W = 768$ is the hidden size. The Wav2Vec2 model is used for feature extraction only and is not tunable during training. For computational efficiency, we reduce the number of timesteps $T$ 8 times using 3 VGG convolutional blocks with sizes of (1, 32), (1, 32), and (2, 32), followed by two fully connected layers to project the previous output hidden size back to model hidden size $D$. The output is then passed through 4 layers of transformer encoder with the same configuration as in the text encoder. As a result, the acoustic encoder transforms the input acoustic features from \(H_{a} \in R^{T \times W}\) to \(H_{a} \in R^{T/8 \times D}\).

The decoder architecture is nearly identical to the text encoder with 6 transformer layers. At each layer, there is an additional cross-attention layer that allows the decoder to focus on the specific parts of the encoder output. To enable the decoder to make use of multiple sources, we apply a serial combination strategy proposed in \cite{combine}. The decoder first attends to the text output as follow:
\begin{equation*}
A_{t} = Attention (Q_{t},K_{t},V_{t}) = Attention (H_{s},H_{t},H_{t})
\end{equation*}
where $Q_{t}$, $K_{t}$, $V_{t}$ denotes the sets of queries, keys, and values. In this case, $Q_{t}$ is the output $H_s$ of the preceding self-attention layer, while $K_{t}$ and $V_{t}$ are copies of the text output $H_{t}$.

To incorporate acoustic features, we utilize the serial combination strategy as follow:
\begin{equation*}
A_{a} = Attention (Q_{a},K_{a},V_{a}) = Attention (A_{t},H_{a},H_{a})
\end{equation*}

\subsection{Training}

All the models are implemented with Hugging Face toolkit \cite{huggingface}. We optimize them using AdamW optimizer (lr = 1e-5, decay = 1e-5) on  NVIDIA V100s. For inference, beam search is used for decoding with a beam size of 8. Our text encoder and decoder are initialized using pre-trained BART-base \cite{bart} (6 layers, 12 attention heads, and hidden size $D$ = 768).

As the text and acoustic features are generated by different encoders, they are dissimilar in terms of representation. This poses a problem as known as heterogeneity in multimodal learning \cite{multimodal}. To homogenize text and acoustic features, we first train the correction model with only the text encoder and decoder. We then replace the text encoder with a newly initialized acoustic encoder and retrain the model on the same dataset. At this step, the decoder's parameters are frozen so that the acoustic encoder is forced to generate acoustic features that are similar to text features. Finally, three pre-trained modules are jointly trained together in which all parameters are tuneable.

\section{Experiment}
\subsection{ASR models}
Two ASR models using Conformer large architecture \cite{conformer} are chosen as the source model for adaptation. The first one is called "GENERAL" as it is trained on 7000 hours of read and conversational speech from various sources. The second one is the "LIBRI" model which is trained on 960h hours from LibriSpeech dataset \cite{libri}. Both models were taken from the NVIDIA NEMO repository \cite{nemo}. For inference, beam search decoding is used in conjunction with an N-gram model to generate transcription hypotheses, which are then re-ranked with a pre-trained GPT2 language model \cite{gpt2}.

\begin{table*}[t]
    \caption{WER(\%) on target domain test set}
    \label{tab:wer_main}
    \centering
    \begin{tabular}{lcccc}
        \toprule
         \ &  \multicolumn{3}{c}{Source model → Target domain } \\
          & GENERAL → NCS & LIBRI → NCS & LIBRI → TED\\
         \midrule
         Base ASR model & 13.3 & 38.1 & 12.4 \\
         \midrule
         Error correction & 10.7 & 30.7 & 10.8 \\
         Self-training adaptation & 10.2 & 20.3 & 9.3 \\
         Self-training $\rightarrow$ \text{Correction} & 9.1 & 18.6 & 8.7 \\
         Self-training with corrected pseudo labels & 9.4 & 18.8 & 8.8 \\
         \bottomrule
    \end{tabular}
\end{table*}

\subsection{Datasets}

We use TED-LIUM 3 (TED) and NCS Part 6 [16] as the target domains for adaptation. The former contains 452 hours of speech extracted from TED conference videos while the latter consists of 1000 hours of speech derived from recorded conversations between two Singapore speakers. We select 5 hours of each dataset for validation and testing. The remaining hours serve as unlabeled data for unsupervised adaptation training.

To generate pseudo training data for error correction, each unlabeled audio is transcribed by an \textit{inferior} and a \textit{superior} ASR model as described in Section \ref{sec:method}. To reduce the distribution mismatch between training and inference, the WER gap between the \textit{inferior} and the \textit{superior} is set so that the WER of the pseudo dataset is equal to the WER of the real validation set. We also remove pseudo samples with WER higher than 0.5 and samples in which the pseudo target has a low confidence score. Finally, we use multiple \textit{inferior} models and apply different rates of Dropout and SpecAugment to generate approximately 1M training samples for each correction model.

\subsection{Results}

To demonstrate the effectiveness of the proposed methods, we train and evaluate our correction models with different adaptation settings. Table \ref{tab:wer_main} shows the corresponding performances in terms of WER on the target domain test set. 

The GENERAL and LIBRI models without adaptation achieved a WER of 13.3\% and 38.1\% on the NCS test set, respectively. This demonstrates a strong domain mismatch between source and target domains, especially on the LIBRI model. However, the mismatch can be reduced significantly by employing a correction model to recover the errors, demonstrated by the WER gains of 2.6\% on GENERAL and 7.4\% on LIBRI. A similar gain of 1.6\% can also be observed when adapting the LIBRI model to the TED domain with correction.

The self-training method presents itself to be a very strong candidate for unsupervised domain adaptation, especially when the domain mismatch is severe. In LIBRI → NCS setting, the method is able to reduce the WER significantly from 38.1\% to 20.3\%. Despite that, the number can be decreased further to 18.6\% by employing a correction model op top of the self-trained model during inference, which is denoted as \textit{\(\text{Self-training} \rightarrow \text{Correction}\)} in Table \ref{tab:wer_main}. However, this practice increases prediction latency. To reap the benefits of error correction while maintaining low latency, we can use the correction model to correct pseudo transcriptions of the unlabeled training data and then proceed with another iteration of self-training. Performances of the final models trained with corrected pseudo labels are shown in the last line in Table 1.

\begin{table}[t]
    \caption{Correction performances in WER (\%) with different input features in GENERAL → NCS adaptation setting}
    \label{tab:features}
    \centering
    \begin{tabular}{p{4.5cm}p{1.0cm}p{1.0cm}}
        \toprule
        & Train & Test\\
         Base ASR model & - & 13.3 \\
        \midrule
         Correction (1-best transcription) & 8.9 & 12.0 \\
         Correction (all features) & 10.0 & 10.7  \\
         \ \ \ \ - w/o dialogue context & 8.4 & 10.9 \\
         \ \ \ \ - w/o acoustic features & 9.1 & 11.8 \\
         \ \ \ \ - w/o text features & 6.6 & 14.1 \\
        \bottomrule
    \end{tabular}
\end{table}

Controlling overfitting is important when training an error correction model with pseudo data. Instead of relying solely on the 1-best transcription, we incorporate dialogue context and acoustic features to not only control overfitting but also to provide additional information. Table \ref{tab:features} shows the impact of different input features in the GENERAL → NCS adaptation setting. As can be seen, the model which utilizes all input features greatly outperforms the one that uses 1-best transcription with an absolute gain of 1.3\% on the test set. Although the dialogue context helps regularize the model, it only brings a small improvement of 0.2\%, which is in contrast to the acoustic features, of 1.1\%. This is because the acoustic features and text features are encoded separately merged afterward. The strategy is equivalent to ensemble learning, which has shown superior performances across many machine learning tasks. However, using acoustic features alone, i.e correction without text features, leads to over-fitting of the pseudo data, at 6.7\%, which translates to a drop in the performance on real test data, at 14.1\%.

Finally, we compare our unsupervised with supervised correction methods for unsupervised domain adaptation. Since labeled data of the target domain is not available, the supervised model can only be trained with transcribed audios of the source domain, expecting that the learned knowledge can be used to correct transcriptions coming from the target domain. For a fair comparison, we choose the LIBRI → TED setting as the domain mismatch is the least severe. Specifically, we first train the supervised model on the LIBRI domain using cross-validation strategy proposed in \cite{nvidia-cor}. The model is then being evaluated and compared to our unsupervised method on the test set of the TED domain. Table \ref{tab:compare} shows the corresponding performances. 

Despite being asked to correct transcriptions coming from a new domain, the supervised model is still able to reduce the WER by 0.4\%. However, the improvement is still far behind compared to the unsupervised one with 1.6\% gain. This can be explained by looking at correction examples in Table \ref{tab:example}. Since the unsupervised model is being exposed to the text distribution of the TED domain, it is able to not only correct domain-specific terms, i.e. \textit{tete → ted}, but also capture domain disfluencies, i.e. \textit{we be → we we}, which does not appear in a read speech domain like LIBRI. On the contrary, the supervised model trained on the LIBRI domain is prone to overfitting, which leads to wrong corrections such as \textit{tete → tad} and \textit{be thought → bethought}, where \textit{tad} and \textit{bethought} appear 264 and 42 times respectively in the LIBRI domain but zero times in the TED domain.

\begin{table}[t]
    \caption{Unsupervised vs supervised error correction performances in WER (\%) in LIBRI → TED adaptation setting }
    \label{tab:compare}
    \centering
    \begin{tabular}{p{3.2cm}p{2.7cm}p{0.7cm}}
        \toprule
         & Source, Label & WER \\
         \midrule
         ASR model & LIBRI, Ground-truth & 12.4 \\
         Supervised correction & LIBRI, Ground-truth & 12.0 \\
         Unsupervised correction & TED, Pseudo & 10.8  \\
        \bottomrule
    \end{tabular}
\end{table}

\begin{table}[t]
    \caption{Correction examples of unsupervised (Unsup-Cor) and supervised correction (Sup-Cor) models}
    \label{tab:example}
    \centering
    \begin{tabular}{p{1.5cm}p{5.0cm}}
        \toprule
         Input & this brings me to my \underline{tete price} wish \\
         Sup-Cor & this brings me to my \underline{tad price} wish \\
         Unsup-Cor & this brings me to my \textbf{ted} \underline{price} wish \\
         Reference & this brings me to my ted prize wish \\
         \midrule
         Input & so we \underline{be} thought that this \\
         Sup-Cor & so we \underline{bethought} that this \\
         Unsup-Cor & so we \textbf{we} thought that this \\
         Reference & so we we thought that this\\
        \bottomrule
    \end{tabular}
\end{table}

\section{Conclusions}
\label{sec:conc}

This study demonstrates how automatic transcriptions generated by ASR models can be improved in domain adaptation scenarios. A novel correction framework was proposed to reconstruct a real transcription from an erroneous one using only unlabeled audios as training data. The method can be applied when one has no control over the ASR systems or can be used to complement existing domain adaptation approaches.

\section{Acknowledgements}
\label{sec:ack}
This work was funded by Science Foundation Ireland through the SFI Centre for Research Training in Machine Learning (18/CRT/6183).

\bibliographystyle{IEEEtran}
\bibliography{mybib}

% \begin{thebibliography}{9}
% \bibitem[1]{Davis80-COP}
%   S.\ B.\ Davis and P.\ Mermelstein,
%   ``Comparison of parametric representation for monosyllabic word recognition in continuously spoken sentences,''
%   \textit{IEEE Transactions on Acoustics, Speech and Signal Processing}, vol.~28, no.~4, pp.~357--366, 1980.
% \bibitem[2]{Rabiner89-ATO}
%   L.\ R.\ Rabiner,
%   ``A tutorial on hidden Markov models and selected applications in speech recognition,''
%   \textit{Proceedings of the IEEE}, vol.~77, no.~2, pp.~257-286, 1989.
% \bibitem[3]{Hastie09-TEO}
%   T.\ Hastie, R.\ Tibshirani, and J.\ Friedman,
%   \textit{The Elements of Statistical Learning -- Data Mining, Inference, and Prediction}.
%   New York: Springer, 2009.
% \bibitem[4]{YourName17-XXX}
%   F.\ Lastname1, F.\ Lastname2, and F.\ Lastname3,
%   ``Title of your INTERSPEECH 2022 publication,''
%   in \textit{Interspeech 2022 -- 23\textsuperscript{rd} Annual Conference of the International Speech Communication Association, September 18-22, Incheon, Korea, Proceedings, Proceedings}, 2022, pp.~100--104.
% \end{thebibliography}

\end{document}